# Pressure-enhanced $f$-electron orbital weighting in $UTe_2$ mapped by quantum interferometry

T. I. Weinberger,[1] Z. Wu,[1] A. J. Hickey,[1] D. E. Graf,[2] G. Li,[3,4]
P. Wang,[3,4] R. Zhou,[3,4] A. Cabala,[5] J. Pu,[6] V. Sechovský,[5]
M. Vališka,[5] G. G. Lonzarich,[1] F. M. Grosche,[1] A. G. Eaton[1*]

[1]Cavendish Laboratory, University of Cambridge,
JJ Thomson Avenue, Cambridge, CB3 0HE, UK

[2]National High Magnetic Field Laboratory, Tallahassee, Florida, 32310, USA

[3]Beijing National Laboratory for Condensed Matter Physics, Institute of Physics,
Chinese Academy of Sciences, Beijing 100190, China

[4]School of Physical Sciences, University of Chinese Academy of Sciences,
Beijing 100190, China

[5]Charles University, Faculty of Mathematics and Physics, Department of
Condensed Matter Physics, Ke Karlovu 5, Prague 2, 121 16, Czech Republic

[6]School of Physics and Astronomy, Shanghai Jiao Tong University, Shanghai, 200240, China

[*]To whom correspondence should be addressed: alex.eaton@phy.cam.ac.uk

**The phase landscape of $UTe_2$ features a remarkable diversity of superconducting phases under applied pressure and magnetic field. Recent quantum oscillation studies at ambient pressure have revealed the quasi-2D Fermi surface of this material. However, the pressure–dependence of the Fermi surface remains an open question. Here we track the evolution of the $UTe_2$ Fermi surface as a function of pressure up to 19.5 kbar by measuring quantum interference oscillations. We find that in sufficient magnetic field to suppress both superconductivity at low pressures and incommensurate antiferromagnetism at higher pressures, the quasi-2D Fermi surface found at ambient pressure smoothly connects to that at 19.5 kbar, with no signs of a reconstruction over this pressure interval. We observe a smooth increase in oscillatory frequency with increasing pressure, indicating that the warping of the cylindrical Fermi sheets continuously increases with pressure. By computing a tight-binding model, we show that this enhanced warping indicates increased $f$-orbital contribution at the Fermi level – up to and beyond the critical pressure at which superconductivity is truncated. These findings highlight the value of high-pressure quantum interference measurements as a new probe of the electronic structure in heavy fermion materials.**

## Introduction

Quantum oscillation (QO) measurements are a powerful direct probe of a material's Fermi surface (FS).[1] The Shubnikov-de Haas (SdH)[2] and de Haas-van Alphen (dHvA)[3] effects measure QOs respectively in the electrical transport and magnetization of metals. These techniques are premised on Landau quantization of itinerant quasiparticles' energy levels in a magnetic field, leading to oscillatory components in derivatives of the free energy (or the density of states) that relate directly to the Fermi surface geometry and carrier effective masses.[4] In sufficiently high magnetic fields magnetic breakdown can occur, whereby quasiparticles tunnel between FS sheets, the detection of which yields information about the spacing of FS sheets in relation to each other.[5–7]

Analogously to the dHvA and SdH effects, in materials with sufficiently close FS sheets for magnetic breakdown to occur in experimentally accessible magnetic field strengths, quantum interference oscillations (QIOs) can be observed in transport measurements at high field.[1,8,9] These stem from interference between quasiparticle orbits that branch into separate paths along the FS before later recombining, typically with one quasiparticle having tunnelled across to another FS sheet and then back again. QIOs thus yield valuable information about how FS sheets connect and span the Brillouin zone. QIOs have been observed in a variety of metals, including elemental magnesium,[9] quasi-2D organic superconductors[10,11] and recently in the heavy fermion superconductor $UTe_2$.[12,13]

In the context of heavy fermion systems, QIO measurements are especially powerful, because compared to dHvA or SdH oscillations they can persist to higher temperatures. This is because the observed frequencies and amplitudes are determined by the *differences* between quasiparticle orbit areas and their effective masses and thus can be observed to much higher temperatures than oscillations stemming directly from Landau quantization.[1,14]

The heavy fermion dichalcogenide $UTe_2$ crystallizes in a body-centred orthorhombic struc-ture ($Immm$ symmetry, space group 71).[15] At ambient pressure and magnetic field it possesses an unconventional superconducting state below a critical temperature $T_c = 2.1$ K, which exhibits numerous characteristics of odd-parity pairing.[16–19] Under the application of either pressure or magnetic field (or both) several other distinct superconducting phases are accessed,[17,19–31] including one that persists to spectacularly high fields in excess of 70 T.[27–31] At a critical pressure of $\approx 15$ kbar superconductivity is abruptly quenched, and an incommensurate antiferromagnetically (AFM) ordered state has been observed at low temperatures.[32] The precise nature of the magnetic properties at high magnetic fields remains the subject of experimental investigation.[33]

The normal state electronic properties of $UTe_2$ at ambient pressure have been probed by angle-resolved photoemission spectroscopy (ARPES) in addition to dHvA and SdH effect measurements,[34–36] which have revealed a remarkably simple FS consisting of two undulating cylindrical sheets, one hole-type and the other electron-type. Slow oscillations (of around 200 T) observed in contactless resistivity measurements by the tunnel diode oscillator (TDO) method in high magnetic fields were reported to be characteristic of an additional, small, 3D FS pocket[12] – but no signature of this pocket was seen in either dHvA[35,36] or SdH[36] effect measurements. However, subsequent high-field measurements reproduced the observation of ref.,[12] but found that the slow oscillations are rapidly suppressed within a 20° rotation away from the crystallographic $a$-axis,[13] inconsistent with a 3D Fermi pocket scenario. Instead, these oscillations can be attributed to QIOs due to their very light (apparent) effective masses – inconsistent with an $f$-electron pocket but very consistent with the close spacing in $k$-space and pronounced undulations of the FS sheets previously revealed by dHvA measurements.[36]

Two recent studies of electrical transport measurements performed on microstructured $UTe_2$ specimens have added further credence to the quasi-2D FS scenario of this material.[37,38] In ref.[37] low frequency QIOs were observed in the contacted resistivity only for magnetic field tilt angles close to the $a$-axis, similar to prior observations in TDO measurements.[13] This is inconsistent with an isotropic dependence on angle that would be expected for a 3D FS scenario.[12] Further corroboration of the quasi-2D nature of the $UTe_2$ FS was provided by directional-dependent resistivity measurements in ref.,[38] which resolved a ~50-fold difference in the low temperature resistivity for current sourced along the direction of the cylindrical axis (the $c$ direction) compared to in the $ab$ plane.

Understanding how the FS of $UTe_2$ evolves under the application of hydrostatic pressure is key to unravelling the rich interplay between magnetic fluctuations, superconductivity, and the underlying heavy fermion physics of this intriguing material.[17] Here, we report a high magnetic field study of the pressure dependence of quantum interference oscillations in $UTe_2$ for magnetic field $\mu_0 H \parallel a$. These QIOs arise from paths that wrap around the $k$-space area enclosed between the cylindrical Fermi sheets (normal to $k_x$).[13] We find that this area continuously increases from 0 to 20 kbar for $\mu_0 H > 15$ T. This means that, for sufficient $H$ to access the paramagnetic normal state above the AFM phase, the $UTe_2$ FS is smoothly connected over this entire pressure interval. Furthermore, we show that this growth in the enclosed area relates to increased warping along the axis of the cylindrical sheets. We present a tight-binding model that accurately reflects the experimentally-determined quasi-2D FS of $UTe_2$, and show that the observed increase in warping of the FS sheets is caused by growing $f$-orbital contribution at the Fermi level under increasing pressure.

## Results

### QIO measurements

Figure 1 shows QIOs in the contactless resistivity of $UTe_2$ for the fixed magnetic field orientation $H \parallel a$ at incremental pressure points up to 19.5 kbar. Data were acquired by the TDO technique[39] (see *Methods*). The frequency of the QIOs continuously increases from 220 T at $p$ = 0.0 kbar up to 330 T at $p$ = 19.5 kbar. We interpret the oscillations for this field orientation as in ref.,[13] where the observation of QIOs in $UTe_2$ was previously discussed. As the frequency of QIOs corresponds directly to the $k$-space area enclosed by the interference paths[40] – sim-ilarly to how dHvA and SdH oscillations are related to enclosed orbital areas by the Onsager relation[1,41] – this indicates that the red shaded area of Fig. 1e has increased in size by a factor of 1.5 from 0 to 19.5 kbar. In our previous analysis, we showed that this area can be directly related to the degree of warping of the electron Fermi surface cylinder (see the Supplementary Information and ref.[13]). Therefore, the increase in area reveals an enhanced warping of the FS cylinders with pressure.

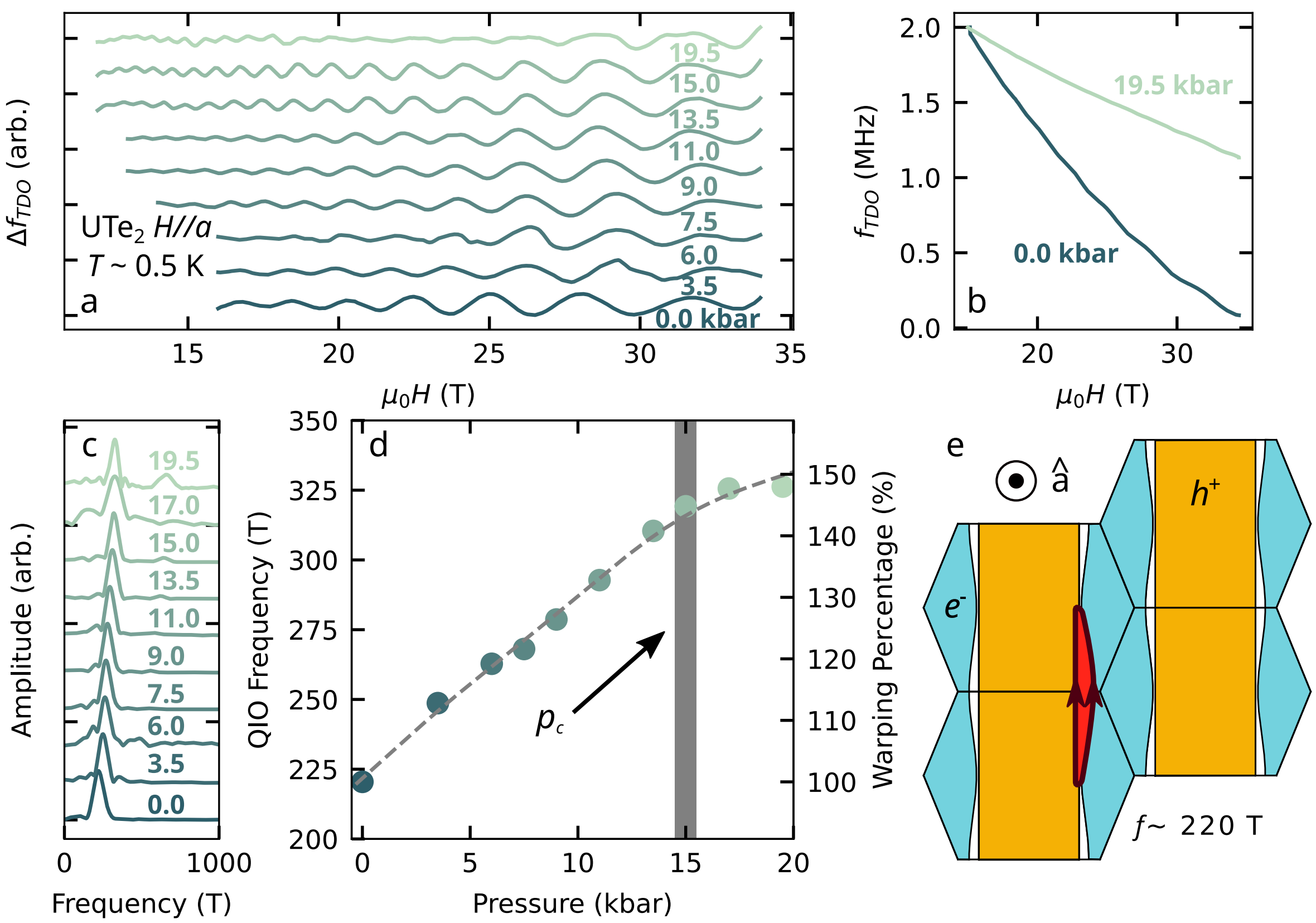

**Fig. 1.** (a) QIOs in the contactless resistivity of $UTe_2$ for $H \parallel a$ measured by the change in resonant frequency of a TDO circuit, $\Delta f_{TDO}$, at various pressures as indicated. Oscillations have been rescaled to be visible on the same scale. (b) Raw contactless resistivity at 0 kbar and 19.5 kbar translated to have the same value at 15 T. The absolute amplitude of the oscillations substantially diminishes from 0.0 kbar to 19.5 kbar. (c) Fast Fourier transforms (FFTs) of the QIOs at each measured pressure point taken over a 17 - 34 T window except the data at 17 kbar where the FFT is taken from 18.5 - 28 T. (d) QIO frequency plotted versus pressure, showing a smooth increase in frequency with compression. The right-hand axis gives the FS warping percentage corresponding to increased QIO frequency. (e) Side-view of the $UTe_2$ FS cylinders (adapted from ref.[36]). The [100] direction (crystallographic $a$-axis) is oriented into the page. The red shaded area corresponds to the enclosed $k$-space area, between the FS cylinders (hole cylinder in orange, electron in blue), which yields a QIO frequency of $\approx$ 220 T.

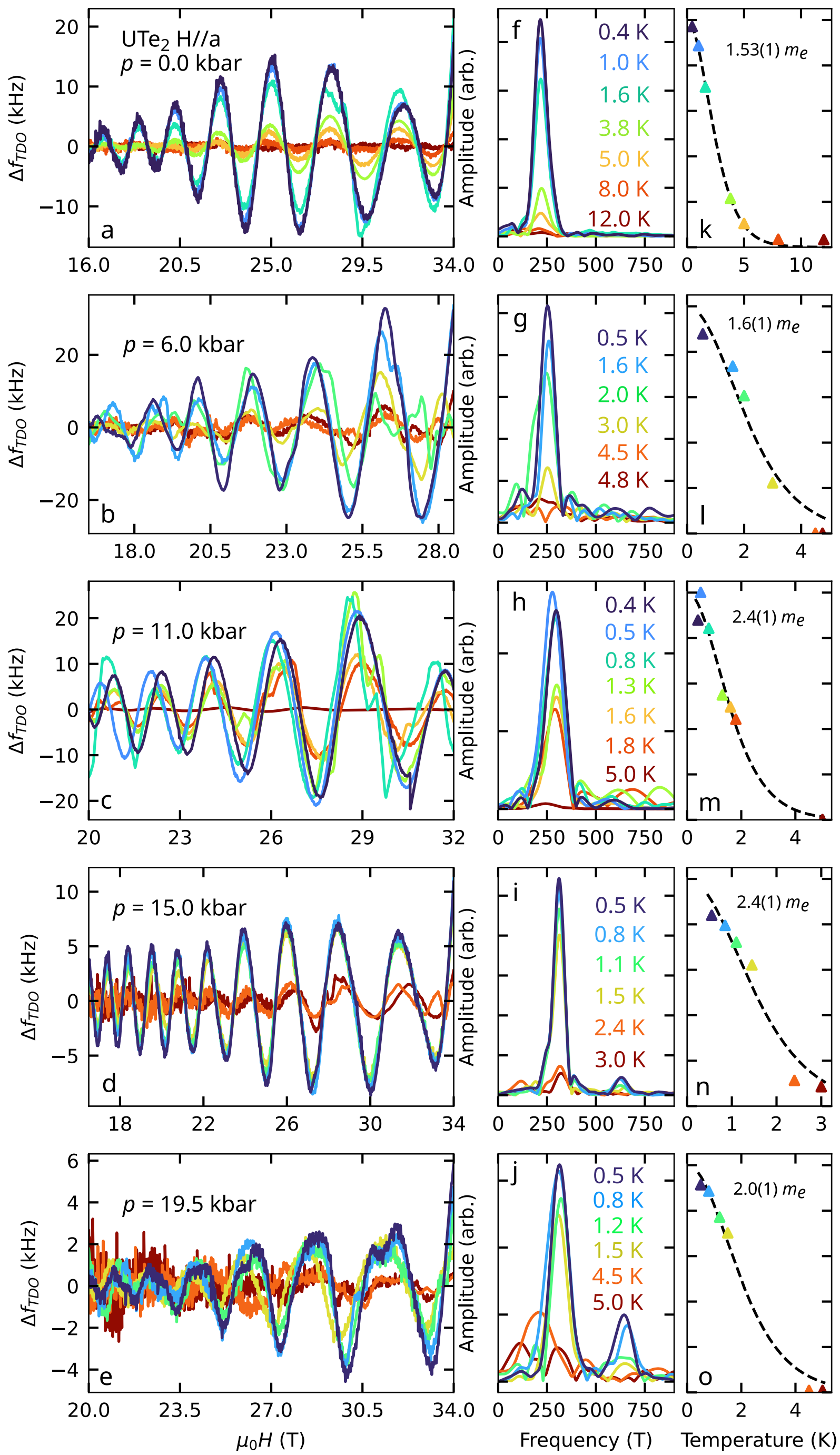

**Fig. 2.** (a-e) QIOs at incremental pressures as labelled, with corresponding fast Fourier transform (FFT) frequency spectra (f-j) and FFT peak amplitudes plotted versus temperature (k-o). A single oscillation with frequency 220-330 T is observed at all pressures, complemented by a second harmonic of increasing amplitude at higher pressures. The colours of the data at each pressure (i.e. of each row of panels) correspond to the temperatures listed by the FFT spectra. Apparent effective masses are extracted from a Lifshitz-Kosevich fit[4] and expressed in terms of the bare electron mass $m_e$.

We measured QIOs for $H \parallel a$ at 10 incremental pressure points, and examined the temperature dependence of the oscillation amplitude at five of these pressures (Figure 2). For each of these pressures, we fit the temperature dependence of the QIO amplitude to the standard Lifshitz-Kosevich formula.[1,4] This yields an apparent effective mass for the QIOs, which reflects the difference in the effective masses of the two quasiparticle trajectories – one along the electron-type FS sheet and the other along the hole-type sheet – which combine to give these QIOs (Fig. 1e). The temperature dependence of the QIO amplitude is given by the derivative of the phase $\phi$ along each interference trajectory with respect to quasiparticle energy $E_k$.[10,42,43] If we express the two trajectories as $\lambda, \lambda'$ we may write the apparent effective mass as $m^*_{\lambda,\lambda'} = \frac{e\hbar\mu_0 H}{2\pi}\left|\frac{\partial(\phi_\lambda - \phi_{\lambda'})}{\partial E_k}\right| = |m^*_\lambda - m^*_{\lambda'}|$ where $e$ is the elementary charge and $\hbar$ the reduced Planck constant.[13,40,42] It is this peculiar property of QIOs – that their apparent effective mass is given by the *difference* between the conventional QO masses for paths $\lambda, \lambda'$ – that enables them to be observed at considerably higher temperatures than dHvA or SdH QOs. This is especially true in a heavy fermion system like $UTe_2$ in which, depending on the magnetic field tilt angle, dHvA experiments have observed effective masses ranging from 32-78 $m_e$.[35,36] In contrast, the QIO apparent effective mass for $H \parallel a$ at ambient pressure is considerably lower at only 1.5 $m_e$. This indicates that the Fermi velocity, $v_F(\mathbf{k})$, is similar over the two arcs that bound the red shaded area in Fig. 1e.

In Figure 3a we plot the magnetic field–pressure phase diagram of $UTe_2$ reported for $H \parallel$

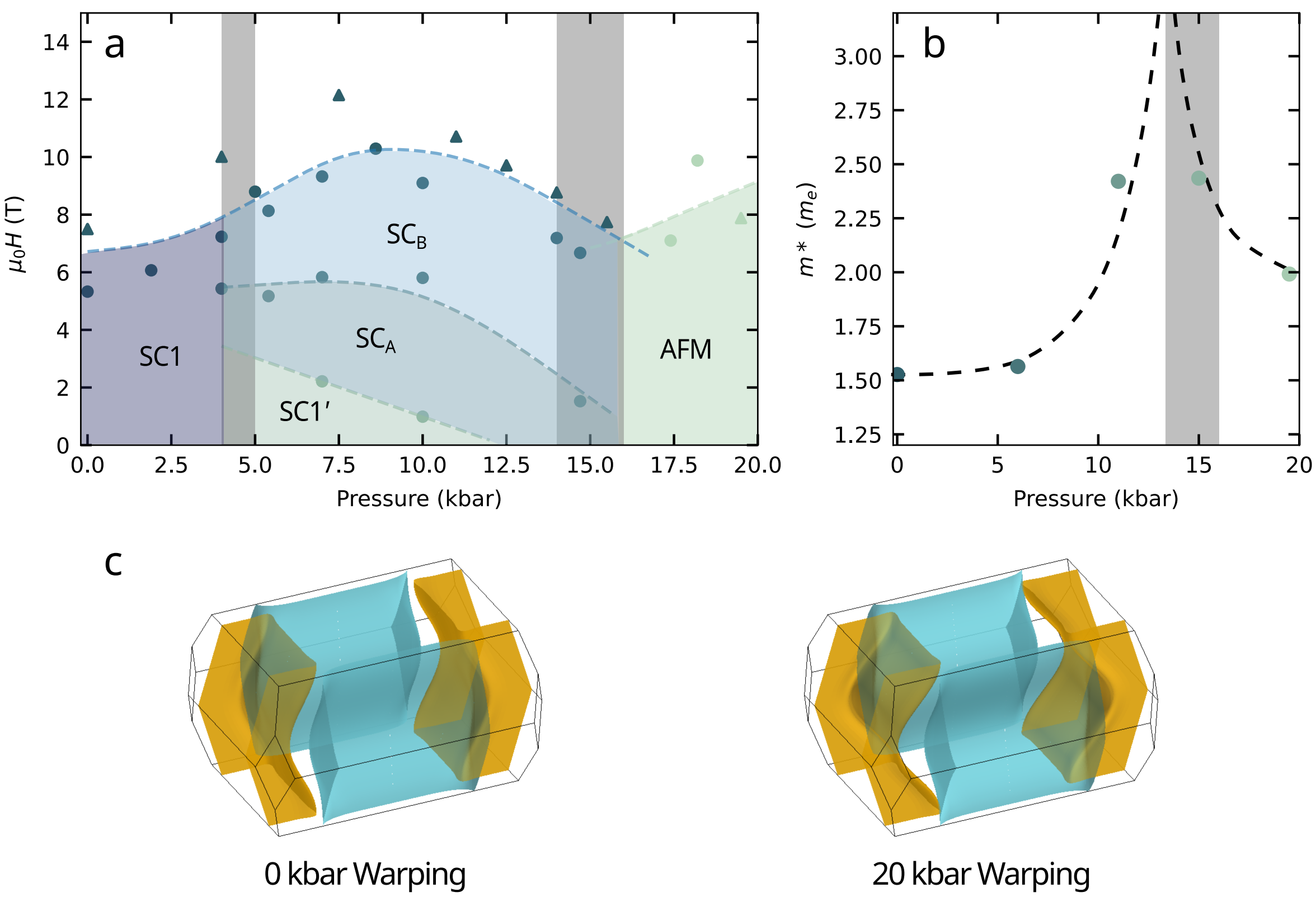

**Fig. 3.** (a) Superconducting phase diagram of $UTe_2$ under hydrostatic pressure for applied magnetic field $H \parallel a$. Circular data points are reproduced from refs.[22,23] in which multiple distinct superconducting states were inferred from specific heat measurements. Triangular points represent superconducting to normal state transitions determined by TDO measurements in this study, with the exception of the highest pressure point that marks the AFM to paramagnetic boundary. Grey shading gives an estimation of the uncertainty in pressure range for the split of SC1 into additional superconducting states, and for the location of the critical pressure. (b) The pressure dependence of the apparent effective masses of $a$-axis QIOs in $UTe_2$ show an enhancement around the critical pressure of $p_c \approx 15$ kbar. (c) Simulations of the degree of warping of the cylindrical Fermi sheets at ambient pressure and 19.5 kbar, which is inferred from the QIO frequency evolution of Fig. 1

$a$,[22,23] along with FS simulations showing the cylindrical warping at ambient pressure and at 20 kbar. Ambient-pressure superconductivity with a single-component order parameter[44,45] has been reported to give way to three additional distinct superconducting phases under pressure for $H \parallel a$.[22,23] How these phases may relate to other superconducting states observed for different magnetic field orientations remains the subject of investigation.[17,19–21,27,28,46–49]

Figure 3b shows the pressure dependence of $m^*$. We observe a strong initial enhancement of $m^*$ with pressure, starting at 1.5 $m_e$ at ambient pressure and reaching a maximum of 2.4 $m_e$ at 15.0 kbar – 60% higher. For further increasing pressure, $m^*$ drops slightly, reaching 2.0 $m_e$ at our highest pressure point of 19.5 kbar and showing a clear peak near $p_c$.

**Fermiology calculations**

The location in $k$-space of the quasiparticle trajectories responsible for the QIOs probed in this study is very well defined (by the red shaded area in Fig. 1e). Therefore, our measurements tell us precisely which sections of the FS sheets undergo a relative change in $m^*$ as a function of pressure. Our experimental results can be interpreted in terms of a six orbital $f$-$d$-$p$ tight-binding calculation (see Methods and Supplementary Information for details) developed according to a similar approach as Ishizuka and Yanase[50] as well as Haruna *et al.*[51] Hopping parameters were modified to fit dHvA measurements[35,36] while also well-representing the calculated bandstructure from GGA+$U$ calculations including spin-orbit coupling.[52] The resulting bandstructure is displayed in Figure 4. The tight-binding model incorporates Te $p$ and U $d$ orbitals that independently produce quasi-1D sheets perpendicular to the $b$ and $a$ directions, respectively, and hybridise to yield a quasi-2D cylindrical Fermi surface as seen in earlier models.[53,54] Our calculation supplements these with U $f$ states just above the Fermi energy, which have been resolved by ARPES measurements.[34] Increasing the hybridisation of the $f$ states with the $p$ and $d$ states mixes $f$-character into the states near the Fermi energy and changes the geometry of the Fermi surface, increasing the degree of warping of the FS cylinders. Figure 5 illustrates how the $f$-orbital contribution at the Fermi level varies as a function of $\mathbf{k}$. Blue regions have low $f$-contribution, whereas red areas possess strong $f$-type character.

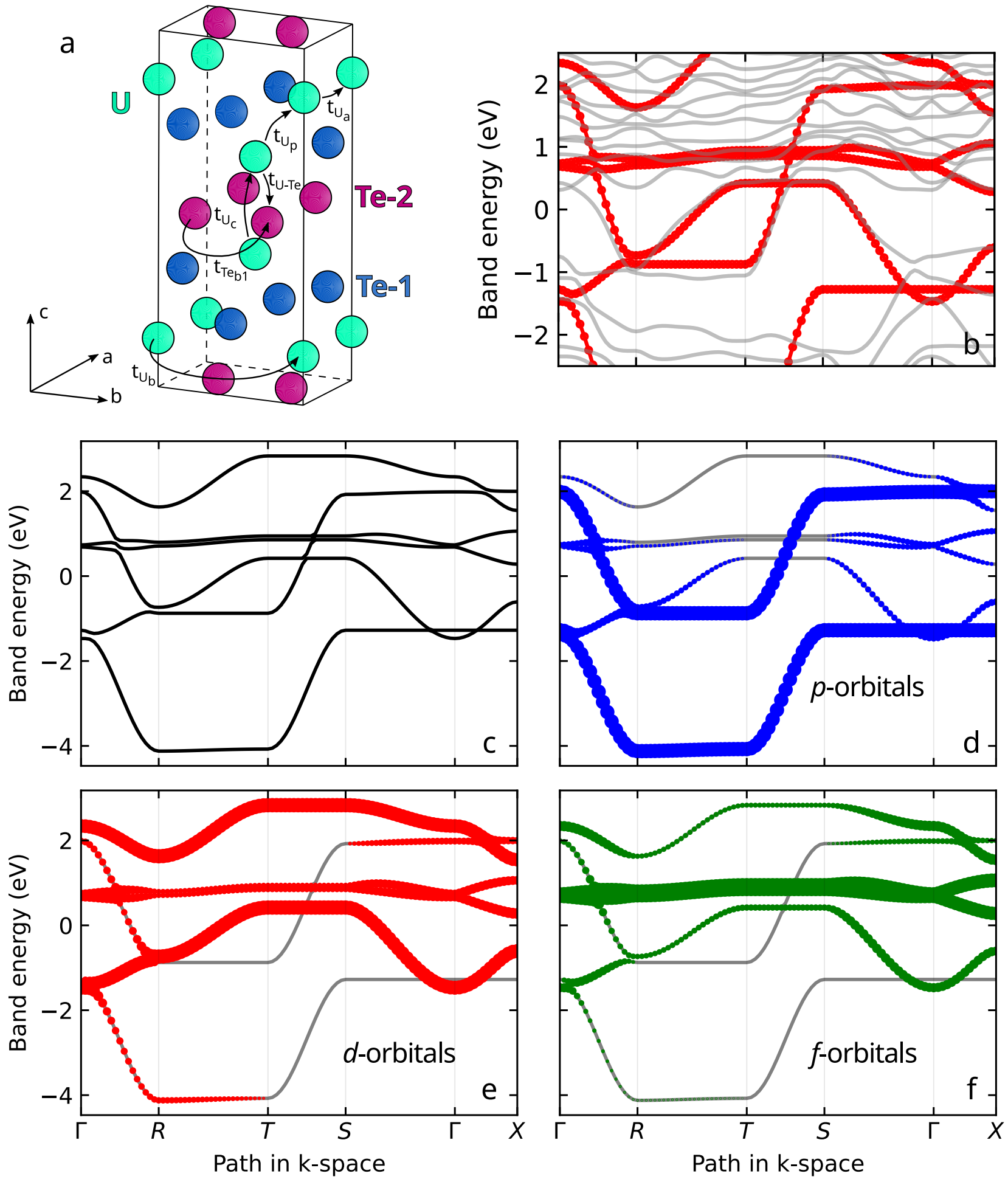

**Fig. 4.** (a) The *Immm* crystal structure of $UTe_2$[17,55] with hopping paths from our tight-binding model (see *Methods* and Supplementary Information). $t_{U_a}$, $t_{U_b}$, and $t_{U_c}$ denote U-U hopping along the $a$-, $b$-, and $c$-directions, respectively, while $t_{U_p}$ represents U-U hopping around a uranium pyramid. The $t_{U-Te}$ term describes hopping between nearest-neighbour uranium and tellurium atoms. Te-Te hopping is primarily along the $b$-direction, with $t_{Te_{b1}}$ accounting for nearest-neighbour hopping and $t_{Te_{b2}}$ for second-nearest-neighbor hopping. $t_{Te_{b2}}$ corresponds to intracell hopping between the equivalent Te $p$-orbital sites. (b) The tight-binding model's band structure (red) compared to a GGA+$U$ calculation with spin-orbit coupling (grey). The $x$-axis denotes the path in $k$-space, as defined in the panels below. (c) The band structure of the tight-binding model and its orbital contributions. Each orbital type features two primary bands. (d) $p$-orbital contributions, represented in blue, are indicated by marker size, with larger markers denoting stronger $p$-character. The higher of the two main $p$-orbital bands contributes to a Fermi surface sheet. (e) $d$-orbital character shown in red. The lower of the two main $d$-character bands also contributes to a Fermi surface sheet. (f) $f$-orbital contributions, displayed in green, reveal that the two main $f$-character bands do not form Fermi surface pockets, as they lie slightly above the Fermi energy. However, significant $f$-character appears at the Fermi level due to strong $f$-$d$ hybridisation. All the bands within this tight-binding model exhibit some degree of mixed-orbital character. However, the lowest energy band is majority Te p-character, and the highest band is majority U d-character. The remaining bands – and notably those that cross the Fermi level – possess a degree of hybridisation between all three orbital characters.

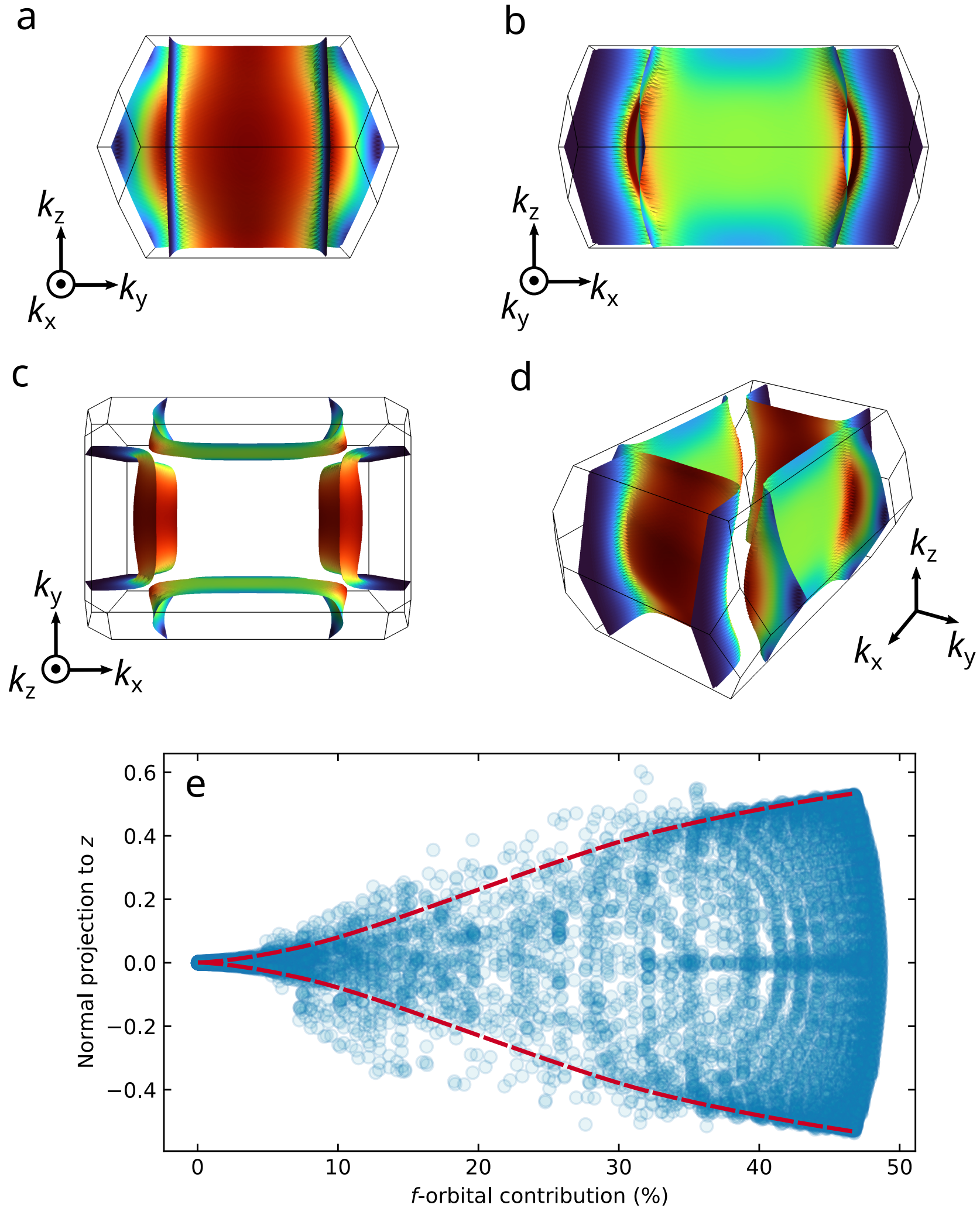

**Fig. 5.** (a-d) The ambient pressure FS of $UTe_2$, constructed from our tight-binding approximation guided by dHvA measurements[35,36] (calculational details given in the Supplementary Information). Red (blue) colouring denotes areas of high (low) $f$-orbital contribution. Areas of higher curvature are found where there is a high $f$-contribution at the Fermi level. (e) Using the $z$-component of a unit vector projected normal to the Fermi surface as a measure of how 3D-like the cylindrical FS warping is, we see that for zero $f$-orbital contribution there is no $z$-component to the FS normal. Therefore, in the absence of $f$ electrons the calculated FS is strictly 2D. Conversely, increasing $f$-electron contribution is associated with an increase in the 3D character of the FS cylinders, caused by an increase in the cylindrical warping.

To capture how the degree of $f$-orbital contribution relates to the warping of the cylindrical FS sheets, we calculated the $k_z$-component of the normal vector to the FS sheets as a function of the $f$-weighting (Fig. 5e). We find that for zero $f$-weighting there is zero projection nor-mal to $k_z$, and therefore the FS would be properly 2D in such a scenario. By contrast, as the $f$-weighting increases, so does the normal projection to $k_z$. In our previous quantum oscillation study we found that the warping of the hole cylinder was more substantial than on the electron cylinder.[36] Consequently, this implies that the hole cylinder exhibits greater $f$-electron contribution than the electron cylinder. Both sheets demonstrate higher $f$-electron contributions on their surfaces adjacent to the $\Gamma$ point. These regions exhibit noticeable warping, while the perpendicular faces remain nearly flat. Although the electron cylinders are only marginally warped, this small degree of warping is sufficient to produce a QIO of approximately 220 T (as shown in the Supplementary Information). Under hydrostatic pressure, the warping of the $p$-electron-dominant faces of the electron cylinders increases, as directly observed through QIO measurements. Indirectly, we infer that this behaviour likely also extends to the $d$-electron-dominant faces of the hole cylinders. However, QIOs cannot be observed for magnetic field oriented along the $b$-axis due to the high critical fields for superconductivity along this direction. Our experimental finding that the warping of the $UTe_2$ FS cylinders smoothly increases under pressure (Fig. 1) is succinctly explained within this model by a continuous increase in $f$-orbital character at the Fermi level.

**Harmonic analysis**

Further information about the evolution of the Fermi surface can be determined by tracking the behaviour of the second harmonic of the fundamental $a$-axis QIO as a function of pressure. Under increasing pressure, we find that the second harmonic grows in amplitude: at ambient pressure, the second harmonic is not visible whereas it is pronounced at 19.5 kbar (Fig. 2j). A second harmonic in the QIO frequency spectrum can occur when, instead of two quasiparticles interfering at the top and bottom of the first Brillouin zone, the quasiparticles start at the bottom of the first Brillouin zone then do not interfere at the top of the first Brillouin zone, but instead only interfere again at the top of the *second* Brillouin zone, such that there is a missed tunnelling event (see Figure 6).

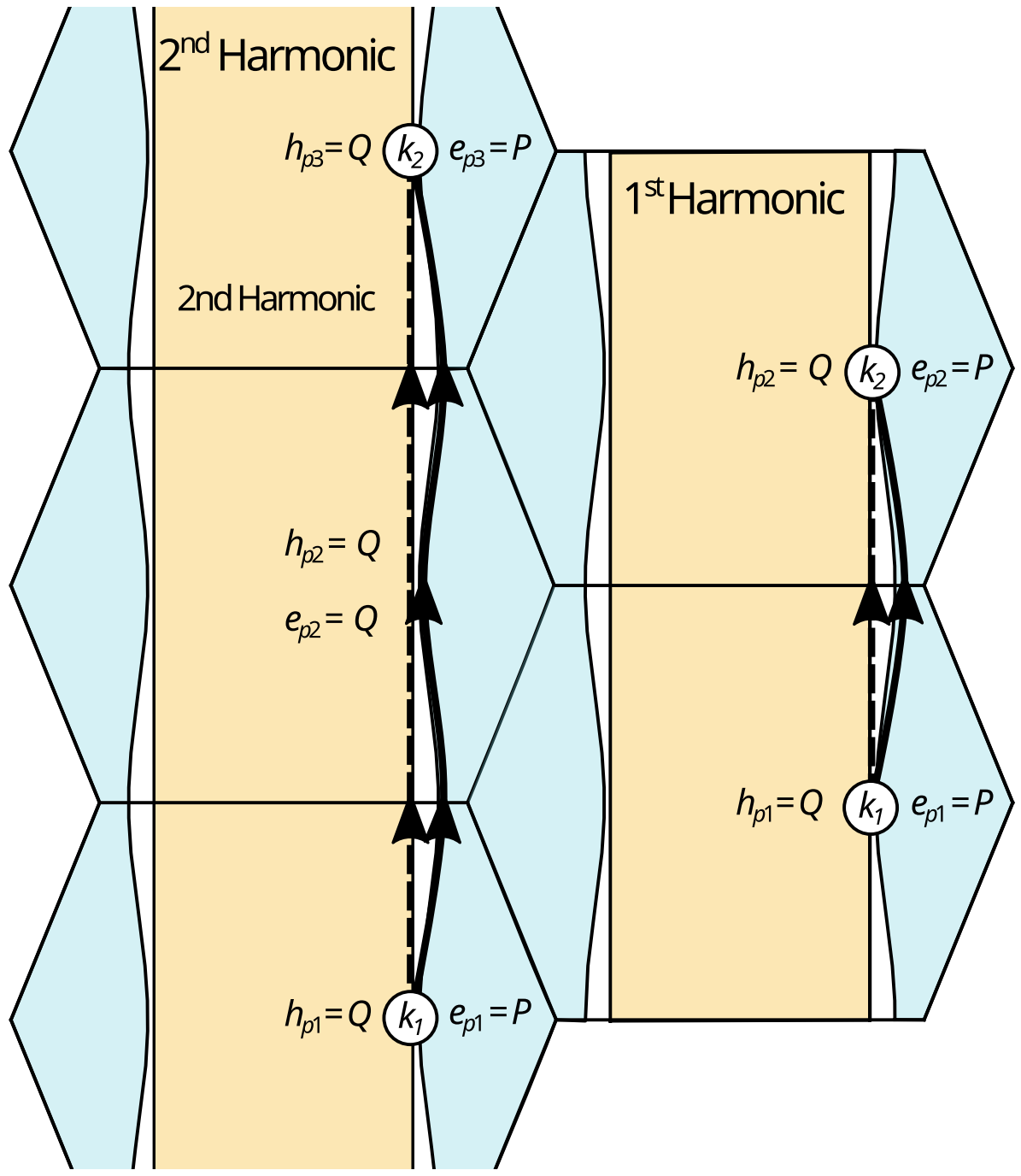

**Fig. 6.** Quasiparticle paths contributing to the first and second harmonic quantum interference oscillations for magnetic field applied along the $a$-axis. Both quasiparticles start and finish at the same point in $k$-space (without loss of generality we will say this is on the hole sheet, coloured ochre, at $k_1$ and $k_2$ respectively). One quasiparticle must then tunnel across to the electron sheet, at which point both quasiparticles must traverse their respective sheets before the quasiparticle on the electron sheet tunnels back onto the hole sheet interfering again at $k_2$. For the second harmonic oscillations to arise, rather than tunnelling back onto the hole sheet at the first instance, the quasiparticle on the electron sheet must skip a tunnelling point ($h_{p2}$ and $e_{p2}$), giving rise to a quantum interference oscillation that incorporates twice the area compared to if it tunnelled directly back at the first instance. The probabilities for tunnelling compared to not tunnelling are $P$ and $Q$ respectively, where $P$ is increasingly favourable if the gap between Fermi surface sheets is small.

If we define a breakdown probability $P$ and the probability of not tunnelling as $Q = 1 - P$, the probability of observing a first harmonic quantum interference oscillation with $H \parallel a$ is $\propto P^2Q^2$: One quasiparticle must stay on its sheet twice with a probability of $Q$ and one must tunnel twice with a probability of $P$. The second harmonic occurs with a probability $\propto P^2Q^4$: one particle must stay on its sheet three times and one must tunnel twice, but when the sheets are close at the centre of the quantum interference arc it must miss a tunnelling opportunity. This means the relative probability of observing a second harmonic quantum interference oscillation compared to the first harmonic goes as $Q^2 = (1-P)^2$. The probability for tunnelling is determined by the probability of magnetic breakdown occurring between the sheets defined as $P = \exp(-B_0/B\cos\theta)$ for a breakdown field $B_0$.[1,7] Since $B_0 \propto k_g^2$, where $k_g$ is the $k$-space gap between sheets,[7,56,57] as the Fermi surface sheets become further apart, the tunnelling probability decreases and so the relative amplitude of the second harmonic quantum interference oscillation should increase relative to the first harmonic. At the same time, the absolute amplitude of the first harmonic should decrease. This is what we observe in $UTe_2$ (see e.g. Fig. 2j and further discussion in the Supplementary Information), indicating that the effect of pressure is to drive the Fermi surface sheets further apart in reciprocal space. This may be due to an increase in the effect of spin-orbit coupling driving increased hybridisation between $p$- and $d$-orbitals.

## Discussion

Prior experimental studies of $UTe_2$ have tracked the evolution under pressure of $A$, the quadratic temperature coefficient of the resistivity $\rho = \rho_0 + AT^2$, where $T$ is temperature and $\rho_0$ is the residual resistivity.[58,59] A clear peak in $A$ has been observed in proximity to $p_c$. As $A^{0.5} \propto m^*$ (where here $m^*$ is strictly the effective [cyclotron] carrier mass,[60] not the apparent QIO mass), this finding was proposed to indicate the presence of an AFM quantum critical point at $p_c$.[58] Our observation of a peak in $m^*$ around $p_c$ provides microscopic evidence in favour of this scenario. In contrast to the peak in the effective mass near $p_c$, the onset of AFM order is not reflected in the QIO frequency, which continues to grow smoothly with increased pressure up to the maximum pressure of 19.5 kbar reached in this study, albeit more gradually above $p_c$ (Fig. 1d).

This indicates that, at least in the high-field paramagnetic normal state in which the QIOs are observed, the FS deforms continuously, with no indication of a sudden reconstruction. Inside the AFM state, the FS may be markedly different, but no oscillatory features could be resolved in our contactless conductivity measurements below the moderately low fields ($\mu_0 H \lessapprox 10$ T) at which the AFM state is suppressed for $H \parallel a$ at $T = 0.4$ K. The peak of $m^*$ in this high field paramagnetic phase is consistent with the expectation of quantum critical behaviour at lower fields.

It has previously been proposed[20] that, for $p > p_c$, the field-polarised paramagnetic state, which is found at ambient pressures for $\mu_0 H_b \gtrapprox 34$ T and which comes down to lower fields at higher pressures,[17,47,61] is also accessed for $H \parallel a$. However, we observed no signatures in our TDO measurements that would signal the metamagnetic transition to the field-polarised phase. Instead, it appears that the field-polarised state is not accessible for $H \parallel a$, at least not for $p \leq 19.5$ kbar with $H \leq 41.5$ T and $T \geq 0.4$ K. The proposal[49] that above $p_c$ the $f$-electrons are largely localized in $UTe_2$ appears to be inconsistent with our findings, at least for $\mu_0 H \gtrsim 10$ T, as our data suggest a continuously increasing $f$-orbital character at the Fermi level under increasing compression up to 19.5 kbar. This raises the question of how the interplay between localized and itinerant properties of the $5f$ electrons[62,63] might evolve in $UTe_2$ at $p > p_c$ from $0 \leq \mu_0 H \lesssim 10$ T – and thus of what role the incommensurate AFM order and its associated magnetic fluctuations may play in forming the various superconductive phases spanning the high-pressure phase landscape.[21,22,47,48,58]

In summary, we tracked key features of the Fermi surface of $UTe_2$ with applied pressure up to 19.5 kbar by measuring quantum interference oscillations (QIOs) using a contactless conductivity technique. We observe a smooth increase in QIO frequency with pressure for magnetic field oriented along the crystallographic $a$-axis. This indicates that the ambient pressure Fermi surface deforms continuously with pressure across the critical pressure, with no evidence of a Fermi surface reconstruction in the high magnetic field paramagnetic state. We show that this deformation is consistent with increasing $f$-orbital contribution at the Fermi level with increasing pressure. We observe a peak in the apparent effective masses of the QIOs around the critical pressure, providing the first microscopic evidence for the presence of quantum criticality underpinning superconductivity in the high-pressure $UTe_2$ phase landscape.

# Methods

### Sample Preparation

Single crystal $UTe_2$ specimens were grown by the molten salt flux (MSF) technique[18] by the procedure given in ref.[36] Samples were screened by a combination of (ambient pressure) resistivity, specific heat capacity, and magnetic susceptibility measurements.

### High Pressure Piston Cylinder Cell Contactless Resistivity Measurements

Samples were oriented using Laue diffractometry and mounted within cylindrical coils of copper wire, with the coil axis aligned parallel to the crystallographic $a$-axis. These coils were connected to copper wires, which were sealed inside tungsten carbide pressure cell feedthroughs using sapphire-loaded Stycast 2850FT epoxy.

The assembled feedthroughs were mounted inside single-walled MP35N pressure cells. Daphne 7575 oil was used as the pressure-transmitting medium,[64] and the pressure at each point was calibrated using the ruby fluorescence method.[65]
We conducted contactless resistivity measurements using the tunnel diode oscillator (TDO) technique.[39] In TDO measurements, the frequency shift $\Delta f$ of an $LC$ circuit is monitored, where the circuit includes a coil coupled to the $UTe_2$ sample with a high effective filling factor $\eta$. As the magnetic field is swept, changes in the resistivity $\rho$ and susceptibility $\chi$ of the sample alter the inductance of the coil, resulting in a frequency shift that may be described by:

$$\frac{\Delta f}{f} \approx -\eta \frac{\delta}{d} \left( \mu_r \frac{\Delta \rho}{\rho} + \Delta \chi \right), \tag{1}$$

where $d$ is the sample thickness, $\mu_r = \chi + 1$, and the skin depth $\delta = \sqrt{\frac{2\rho}{\mu_r \mu_0 \omega}}$, with excitation frequency $\omega$. In good metals like $UTe_2$, where $\rho$ is low and $\delta$ is short, $\Delta f$ is predominantly sensitive to changes in $\rho$. Consequently, TDO is well-suited for contactless resistivity measurements in high-conductivity materials.

TDO measurements typically offer higher resolution compared to alternative contactless methods for measuring changes in the resistivity such as the proximity detector oscillator (PDO)[66,67] technique. However, they require a low line impedance between the measurement coil and the tunnel diode.[39] This necessitates placing certain electronic components, such as the tunnel diode and associated circuitry, within the cryostat near the sample. For this study, we employed a TDO setup in steady (dc) magnetic fields, following protocols similar to those outlined in ref.[68]

Experiments were conducted in a resistive magnet up to 41.5 T fitted with a $^3$He cryostat at the National High Magnetic Field Lab, Florida, USA; and in a superconducting magnet up to 30 T with a dilution refrigerator at the Synergetic Extreme Condition User Facility, Chinese Academy of Sciences, Beijing, China.

### Lifshitz-Kosevich Analysis of Apparent Quantum Interference Effective Masses

Fig. 2 presents quantum interference oscillations (QIOs) measured across varying temperatures, with a pronounced reduction in oscillatory amplitude at higher temperatures. The *apparent* effective mass associated with QIOs reflects the mass difference between the quasiparticle orbits involved. For interfering trajectories $\lambda$ and $\lambda'$, the effective mass difference can be expressed as:

$$m^{\lambda,\lambda'} = \frac{e\hbar\mu_0 H}{2\pi}\left|\frac{\partial\left(\phi_\lambda - \phi_{\lambda'}\right)}{\partial E_k}\right| = |m_\lambda - m^*_{\lambda'}|, \tag{2}$$

where $e$ is the elementary charge and $\hbar$ is the reduced Planck constant.[13] The apparent QIO mass, $m^*$, is determined by fitting the temperature dependence of the FFT amplitudes to the Lifshitz-Kosevich (LK) temperature damping formula. This fit is depicted in Fig. 2k-o.

The temperature damping factor, $R_T$, follows the LK expression:[1]

$$R_T = \frac{X}{\sinh X}, \tag{3}$$

where

$$X = \frac{2\pi^2 k_\mathrm{B} T m^*}{e\hbar B}, \tag{4}$$

with $k_\mathrm{B}$ as Boltzmann's constant, $T$ as temperature, and $B$ as the mean magnetic field strength over the inverse field range used in the FFT computation. The apparent effective mass $m^*$ is extracted by fitting the QIO amplitude to Eqn. 3 as a function of temperature.

### Tight-Binding Model

Our tight-binding model incorporates nearest-neighbour hopping between uranium atoms along the $a$-, $b$-, and $c$-directions, as well as across the pyramidal uranium arrangements at the top and bottom of the conventional unit cell. For Te-Te $p$-orbital hopping, we consider only the $b$-direction, and U-Te hopping is restricted to nearest neighbours. Studies suggest that U $d$- and Te $p$-orbitals interact via a Rashba-like antisymmetric spin-orbit coupling mechanism,[50,51,54] which defines the cylindrical Fermi surface geometry.

We set the on-site potential of the Te $p$-orbitals below the Fermi level, while $d$-orbitals lie above it. The relative positions of these orbitals govern the Fermi surface cylinder dimensions. Narrow $f$-orbital bands, located just above the Fermi level between the $d$- and $p$-orbitals, do not cross the Fermi level. Fermi surface warping is tunable via the U-U $f$-orbital hopping around Uranium pyramids and the interaction strength between U $f$- and $d$-orbitals. These parameters are calibrated to fit quantum oscillation data (Figure S2, Table 1)[36] while still recreating the bandstructure calculated using GGA+$U$[52] including spin-orbit coupling. Eigenvalues and eigenvectors were evaluated using the PythTB package.[69]

| | Atom | $ff$ | $fd$ | $fp$ | $dd$ | $dp$ | $pp$ |
|---|---|---|---|---|---|---|---|
| $t_{U_a}$ | U-U | -0.0375 | -0.025 | - | -0.3 | - | - |
| $t_{U_b}$ | U-U | - | - | - | - | - | - |
| $t_{U_c}$ | U-U | -0.05 | -0.075 | - | -1.2 | - | - |
| $t_{U_p}$ | U-U | -0.05 | -0.15 | - | -0.1 | - | - |
| $t_{U-Te}$ | U-Te | - | - | 0.005 | - | -0.1 | - |
| $t_{Te_{b1}}$ | Te-Te | - | - | - | - | - | 1.6 |
| $t_{Te_{b2}}$ | Te-Te | - | - | - | - | - | 0.7 |

| | Atom | $\epsilon$ |
|---|---|---|
| $\epsilon_{U_f}$ | U | 0.825 |
| $\epsilon_{U_d}$ | U | 1.025 |
| $\epsilon_{Te_p}$ | Te | -1.075 |

**Table 1:** The tight-binding hopping parameters $t_{ij_x}$ describe hopping from an orbital of type $i$ to one of type $j$ along direction $x$, where $i, j \in [f, d, p]$ and $x \in [a, b, c]$. The on-site potential for the $i^{\text{th}}$ orbital is denoted by $\epsilon_i$. Initial parameters were based on those reported by Ishizuka *et al.*[50] and iteratively refined to better fit quantum oscillation data (Figure S2). The final parameters were rescaled to match the bandstructure from GGA+$U$ calculations including spin-orbit coupling (Fig. **??**).[52] All values are expressed in electron volts (eV)

| Species | $x$ | $y$ | $z$ |
|---|---|---|---|
| U | 0 | 0 | 0.1367 |
| Te2 | 0 | 0.2532 | 0.5 |

**Table 2:** The internal atomic positions used in our tight-binding model of $UTe_2$. The lattice constants used were $a$ = 4.16 Å, $b$ = 6.13 Å, and $c$ = 13.96 Å.

# Acknowledgements

We are grateful to D.V. Chichinadze, D. Shaffer, J. Schmalian, T. Hazra, A. McCollam, N.R. Cooper, P. Coleman, A.F. Bangura and especially A. Carrington and T. Helm for stimulating discussions. We thank T.J. Brumm for technical assistance. This project was supported by the EPSRC of the UK (grants EP/X011992/1 & EP/R513180/1). A portion of this work was performed at the National High Magnetic Field Laboratory, which is supported by National Science Foundation Cooperative Agreement No. DMR-2128556 and the State of Florida. Crystal growth and characterization were performed in MGML (mgml.eu), which is supported within the program of Czech Research Infrastructures (project no. LM2023065). We acknowledge financial support by the Czech Science Foundation (GACR), project No. 22-22322S. A portion of this work was carried out at the Synergetic Extreme Condition User Facility (SECUF). T.I.W. and A.G.E. acknowledge support from QuantEmX grants from ICAM and the Gordon and Betty Moore Foundation through Grants GBMF5305 & GBMF9616 and from the US National Science Foundation (NSF) Grant Number 2201516 under the Accelnet program of Office of International Science and Engineering (OISE). Computational simulations were performed using the ARCHER2 UK National Supercomputing Service (https://www.archer2.ac.uk). A.G.E. acknowledges support from the Henry Royce Institute for Advanced Materials through the Equipment Access Scheme enabling access to the Advanced Materials Characterisation Suite at Cambridge, grants EP/P024947/1, EP/M000524/1 & EP/R00661X/1; and from Sidney Sussex College (University of Cambridge).

# Competing interests statement

The authors declare no competing interests.

# Supplementary information for:

## Pressure-enhanced $f$-electron orbital weighting in $UTe_2$ mapped by quantum interferometry

T. I. Weinberger, Z. Wu, A. J. Hickey, D. E. Graf, G. Li, P. Wang, R. Zhou, A. Cabala, J. Pu, V. Sechovský, M. Vališka, G. G. Lonzarich, F. M. Grosche, A. G. Eaton

Correspondence to: alex.eaton@phy.cam.ac.uk

## 1 Relationship between Quantum Interference Oscillation Frequencies and Fermi Surface Warping

The quantum interference oscillation (QIO) frequency observed for magnetic field oriented parallel to the $a$-axis of $UTe_2$ can be related to the degree of warping of the electron pocket of the Fermi surface in our parameterized Fermi surface model.[1] The 220 T QIO arises from quasiparticles moving along $k_z$ on the electron and hole sheets (see Fig.1e in the main text). These oscillations correspond to the warping of the electron cylinder in the $k_y$ direction as it deviates away from, and returns towards, the hole cylinder. While the hole cylinder follows a straight vertical trajectory in $k_z$ when viewed along the $a$-axis, the electron cylinder traces a sinusoidal path in $k_y$, described by

$$d_y = r_e\left(1 - \cos(k_z)\right), \tag{1}$$

where $r_e$ is the warping amplitude. In our ambient pressure Fermi surface model, we determined this warping parameter to be $0.006a_0^{-1}$ (with $a_0$ as the Bohr radius), based on fits to our Fermi surface model.[1,2] The area of the resulting loop is given by

$$A = \int_{-\pi}^{\pi} r_e \left(1 - \cos(k_z)\right) dk_z = 2\pi r_e. \tag{2}$$

Our parameterised model yields an enclosed area corresponding to a QIO frequency of 207 T at 0 kbar, in close agreement to the measured value of 220 T.

# 2 Probability Dependence of Quantum Interference Oscillations and Harmonics

As discussed in the main text, the probability dependence of the first harmonic of a QIO for fields along the $a$-axis is $P_{QIO_1} \propto P^2Q^2$ where $P$ is the probability of a breakdown event and $Q = 1 - P$ is the probability of avoiding a breakdown event. To observe the second harmonic, two further breakdown events must be avoided such that $P_{QIO_2} \propto P^2Q^4$.

This gives rise to the interesting property that although as $P \longrightarrow 0$ (i.e. there is no breakdown) both $P_{QIO_1} \longrightarrow 0$ and $P_{QIO_2} \longrightarrow 0$, the relative amplitude of the second harmonic to the first goes $P_{QIO_2}/P_{QIO_1} \longrightarrow 1$ (Fig. S1a). While a full treatment would also consider higher harmonics, for simplicity here we shall restrict our considerations to the first two. The breakdown probability depends on the breakdown field $B_0$:

$$P = \exp(-B_0/B\cos\theta). \tag{3}$$

The breakdown field is proportional to the square of the $k$-space gap between Fermi surface sheets, $B_0 \propto k_g^2$ (ref.[3]). As $k_g$ increases, $P_{QIO_2}/P_{QIO_1}$ approaches 1. We observe that as we increase the pressure under which our $UTe_2$ sample is subjected, the ratio of the second harmonic to the first harmonic increases, which we interpret as indicating that the gap between the Fermi surface sheets grows (Fig. S1b).

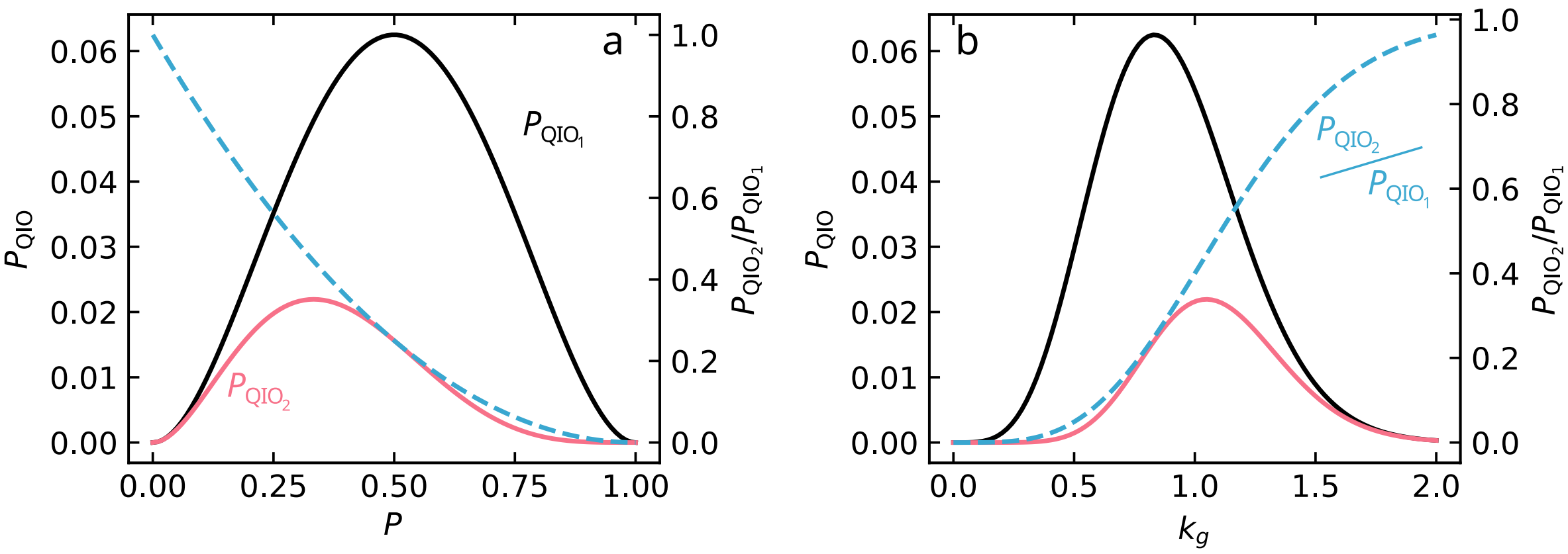

**Fig. S1.** (a) Probability of the first harmonic QIO (black) and the second harmonic QIO (red) occurring as a function of breakdown probability $P$ (indicative values from the left-hand axis). Since both breakdown events and avoided breakdown events are required for a QIO to occur, the QIO probability does not monotonically increase with increased breakdown probability. Instead, it exhibits a well-defined peak in the range $P \in [0, 1]$. Interestingly, the ratio of the second harmonic to the first harmonic, $P_{QIO_2}/P_{QIO_1}$ (dashed blue), increases as the probability of breakdown decreases (indicative value on the right-hand axis). (b) $P$ can be related to the $k$-space gap between Fermi surface sheets by the functional form $P \approx \exp(-k_g^2)$. This shows that as $k_g \to \infty$, $P_{QIO_2}/P_{QIO_1} \to 1$. Our data show that $P_{QIO_2}/P_{QIO_1}$ grows under pressure. Consequently, we interpret that as indicating the gap between Fermi sheets grows as pressure is increased.

# 3 Comparison of Tight-Binding and GGA+$U$ Band Structures

To investigate the orbital contributions at the Fermi level, we developed a minimal six-orbital tight-binding model. Since $UTe_2$ adopts an inversion-symmetric *Immm* crystal structure and is paramagnetic at ambient pressure and zero magnetic field, spin polarisation is neglected. The model assumes that the Fermi surface is primarily influenced by U $f$- and $d$-orbitals and Te $p$-orbitals. The unit cell contains two uranium atoms (each contributing $f$- and $d$-orbitals) and four tellurium atoms (with only the Te-2 atoms contributing $p$-orbitals at the Fermi level). This $f$-$d$-$p$ model is consistent with previous studies.[4,5] Initial hopping parameters were based on Ishizuka *et al.*[4] and iterated to match our quantum oscillation data (Figure S2).

The geometry of the Fermi surface is insensitive to the absolute values of the hopping parameters, provided they are all scaled equally. Therefore, to confirm the validity of our model and rescale the parameters, we compared our tight-binding model to a density functional theory (DFT) calculation for $UTe_2$ with internal coordinates shifted to produce an improved fit to our quantum oscillation data (Figure 4).

We performed DFT calculations for $UTe_2$ using the full-electron, linearised augmented plane-wave package Wien2K.[6] The electronic structures were converged on a $17 \times 17 \times 17$ Monkhorst-Pack $k$-mesh within the Brillouin zone of the primitive unit cell using the Generalized Gradient Approximation exchange-correlation potential. A Hubbard parameter, $U$, of 8 eV was applied, with the static magnetic moment on uranium ions constrained to zero. Spin-orbit coupling effects were also included. Lattice parameters and atomic positions were initialised based on prior DFT studies[7,8] and then perturbed to better fit quantum oscillation data of $UTe_2$ (Table S1).

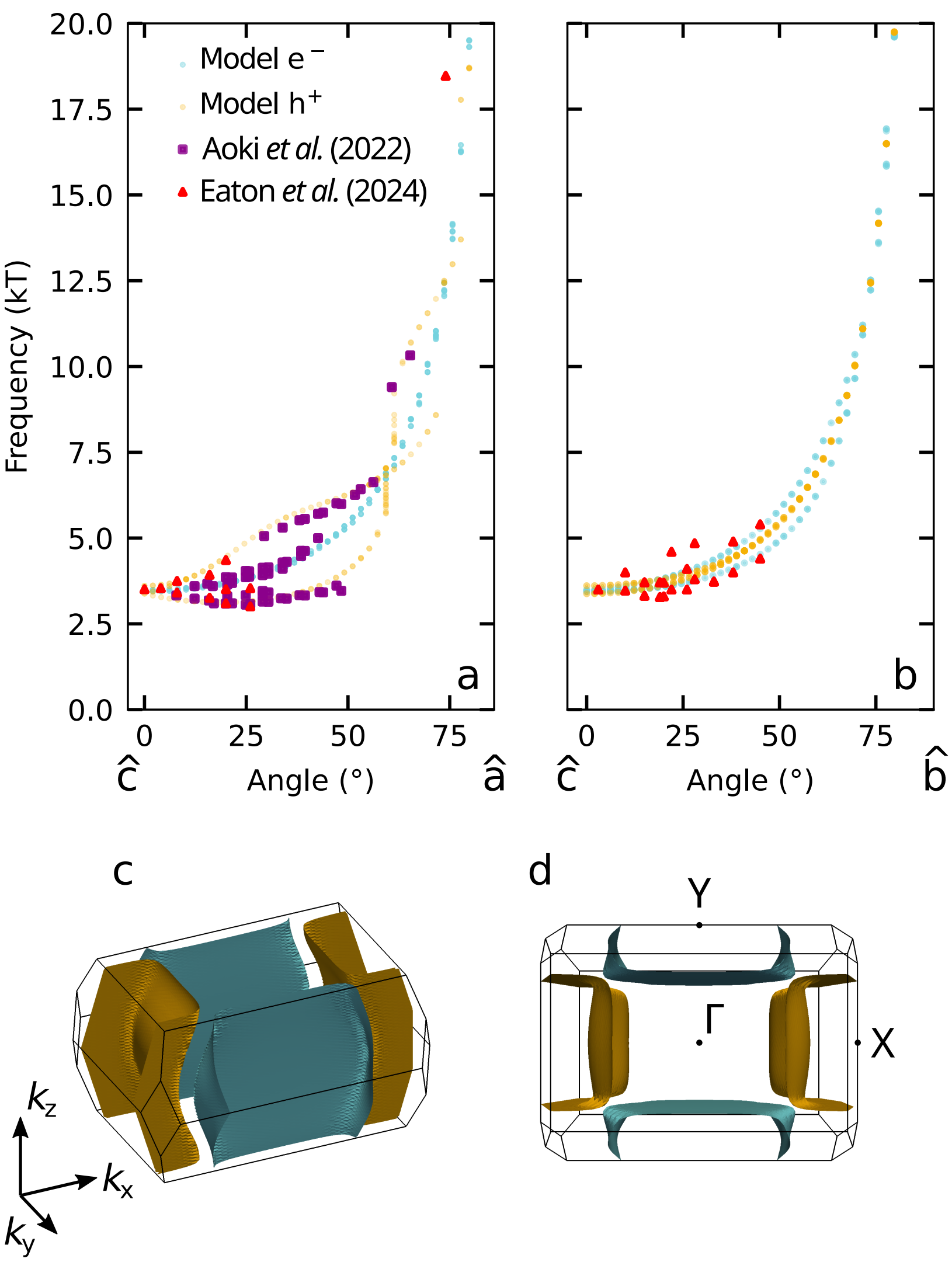

**Fig. S2.** The Fermi surface of the tight-binding model and its fit to quantum oscillation data are shown for (a) the $c$-$a$ magnetic field tilt plane and (b) the $c$-$b$ plane.[1,8] Pronounced warping of the hole sheet is evident in (c) the side view of the Fermi surface, showing the undulating warping of the Fermi sheets, also visible from (d) the top-down view.

| Species | $x$ | $y$ | $z$ |
|---|---|---|---|
| U | 0 | 0 | 0.13473 |
| Te1 | 0.5 | 0 | 0.26299 |
| Te2 | 0 | 0.21438 | 0.5 |

**Table S1:** The internal atomic positions used in our DFT study of $UTe_2$. The lattice constants chosen for the calculations were $a$ = 4.16 Å, $b$ = 6.13 Å, and $c$ = 13.96 Å.

# 4 Orbital Weighting of $f$-, $d$-, and $p$-orbitals

The orbital weighting of the tight-binding model's bands and Fermi surface highlights the relationship between Fermi surface geometry and orbital character. The two bands with the strongest $f$-orbital character exhibit narrow bandwidths, consistent with heavy, localized $f$-electron behaviour. These $f$-bands do not cross the Fermi level, resulting in no additional 3D pockets at the Γ or Z points. However, significant $f$-$d$ hybridisation in the $d$-dominated bands (Figure 4) explains the heavy quasiparticle masses observed by quantum oscillation measurements and the $f$-orbitals' influence on electronic and superconducting properties, even without forming Fermi sheets.

The $d$-bands, split around the Fermi level, have wider bandwidths and lower effective masses compared to $f$-bands, with the lower $d$-band crossing the Fermi level. In contrast, $p$-bands are also split about the Fermi level, with the upper band contributing to an electron-like Fermi surface sheet. The $p$-bands show limited $f$- and $d$-character except between the Γ and R high-symmetry points.

The tight-binding model accurately reproduces the Fermi surface's orbital contributions (Figure S3). Fermi surface sheets perpendicular to the $k_y$-direction are primarily $p$-dominated, while those perpendicular to the $k_x$-direction are $d$-dominated. The $f$-character is most prominent on the warped face of the hole sheet perpendicular to $k_x$-direction, aligning with the quasi-2D behaviour of the Fermi surface and suggesting that $f$-electrons drive this effect. This could have important consequences for the nature of spin fluctuations in $UTe_2$.

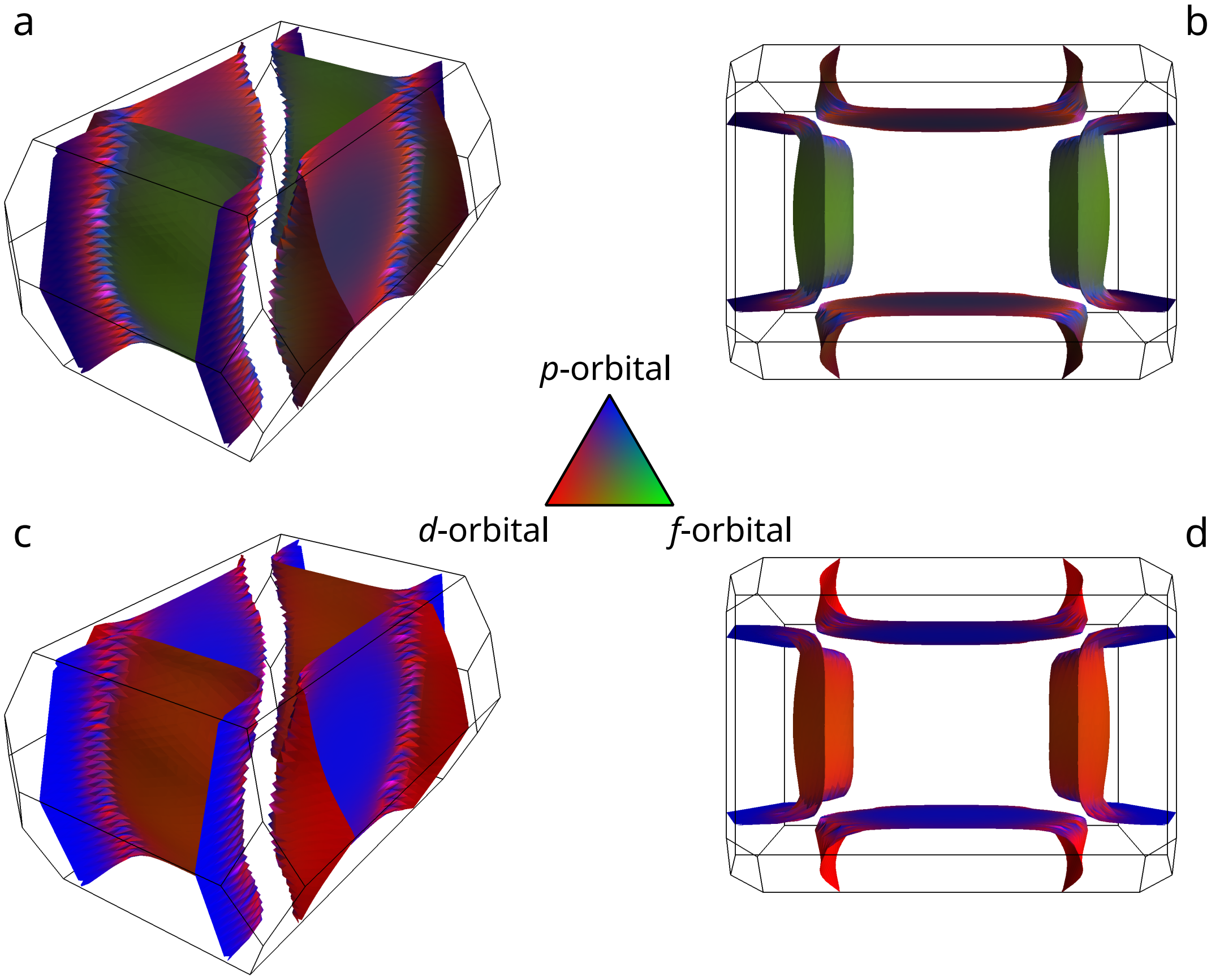

**Fig. S3.** (a, b) Orbital contributions of $p$- (blue), $d$- (red), and $f$-orbitals (green) to the Fermi surface sheets. The $f$-orbitals predominantly contribute to the warped face of the hole cylinder oriented perpendicular to the $k_x$-direction. (c, d) The squared orbital contributions reveal that the Fermi surface is primarily of $p$- and $d$-character, with $p$-orbitals dominating the character of FS sheets perpendicular to the $k_y$-direction and $d$-orbitals perpendicular to the $k_x$-direction.